\documentclass{article}
\pdfoutput=1
\usepackage{arxiv}

\usepackage[utf8]{inputenc} 
\usepackage[T1]{fontenc}    
\usepackage{hyperref}       
\usepackage{url}            
\usepackage{booktabs}       
\usepackage{amsfonts}       
\usepackage{nicefrac}       
\usepackage{microtype}      
\usepackage{lipsum}

\usepackage{amsmath}
\usepackage{graphicx,psfrag,epsf}
\usepackage{enumerate}

\title{\bf Domain of Inverse Double Arcsine Transformation}

\author{Jong-Hyeon Jeong \\
	Department of Biostatistics\\
	University of Pittsburgh\\
	Pittsburgh, PA 15216\\
	\texttt{jjeong@pitt.edu}
	}

\begin{document}
	\maketitle

\begin{abstract}
	To combine the proportions from different studies for meta-analysis, Freeman and Tukey double arcsine tranformation can be useful for normalization and variance stabilization. The inverse function of the double arcsine transformation has been also derived in the literature to recover the original scale of the proportion after aggregation. In this brief note, we present the domain and range of the inverse double arcsine transformation both analytically and graphically. We notice an erratic behavior in the mathematical formula for the inverse double arcsine tranformation at both limits of its domain, and propose approximation methods for both small and large samples. We also propose a simple accuracy measure, the maximum percent error (MPE), of the large sample approximation, which can be used to determine the sample size that would provide a certain accuracy level, and conversely to determine the accuracy level of the approximation given a sample size.
\end{abstract}

\section{Introduction}

The meta-analysis is often performed to aggregate the proportions over multiple studies. However, the event rates can be null from some studies yet they still need to be included in the analysis to represent the whole population. In such cases, the resulting distribution of the proportions  tends to be 0 inflated and for the purpose of normalization and variance stabilization, the Freeman-Tukey double arcsine transformation (Freeman and Tukey, 1958) has been popular. Mosteller and Youtz (1961) provided Tables of this tranformation. 

The aggregated number of the transformed values from a meta-analysis needs to be transformed back to the original scale of the proportion for easier interpretation. Miller (1978) presented the inverse function of the double arcsine tranformation, but not much attention or details were given to the domain of its inverse transformation. Barendregt {\it et al.} (2013) pointed out that the inverse arcsine transformation is numerically unstable only near 0 in its domain and proposed an approximation method, apparently based on simulation studies without mathematical justification. In this note, however, we show that the inverse transformation function has an erratic behavior in both limits of the domain, i.e. both near 0 and near 1, and provide approximation methods for both small and large samples. Unlike Barendregt {\it et al.} (2013) where a fixed constant was used for the domain of the inverse transformation for approximation near 0, we derive a flexible accuracy measure of the large sample approximation that is a function of the sample size.

\section{Domain and Range of Double Arcsine Transformation and its Inverse}
Suppose $x$ and $n$ denote the number of cases (successes) and the number of subjects (trials). The double arcsine transformation is defined as (Freeman and Tukey, 1950) 
\begin{equation}
\theta(p)=(1/2)\left\{\sin^{-1}\left(\sqrt{p/(1+1/n)}\right)+\sin^{-1}\left(\sqrt{(p+1/n)/(1+1/n)}\right)\right\}, \quad p \in [0,1]\label{eqn;1}
\end{equation}
where $p=x/n$. We can immediately see that double arcsine transformation reduces to the simple arcsine transformation as $n \rightarrow \infty$, i.e.
\begin{equation}
\lim_{n \rightarrow \infty}\theta(p)=\sin^{-1}(\sqrt{p}), \quad p \in [0,1].\label{eqn;2}
\end{equation}
Note that the range of double arcsine transformation is from $\theta(0)=(1/2)\sin^{-1}\left(\sqrt{1/(n+1)}\right)$ to $\theta(1)=\pi/4+(1/2)\sin^{-1}\left(\sqrt{n/(n+1)}\right)$, and as $n \rightarrow \infty$ the range is extended to $[0,\pi/2]$. Figure 1 plots the double arcsine transformation in equation (\ref{eqn;1}) for different $n$'s, where $\sin^{-1}(\sqrt{p})$ is the limiting function as $n \rightarrow \infty$. 
\begin{figure}
	\centering
	\includegraphics[width=0.8\linewidth]{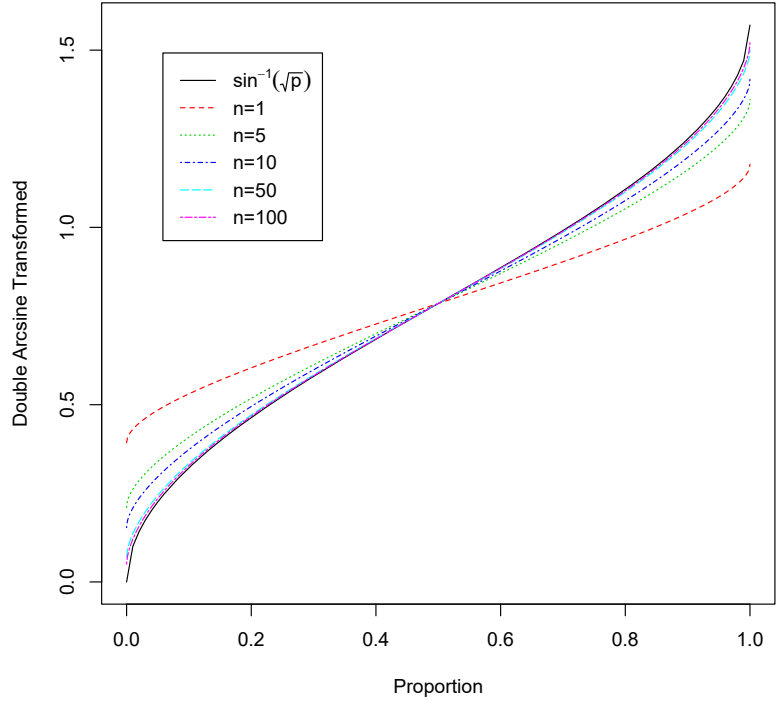}
	\caption{Graph of double arcsine transformation for different sample sizes; $\sin^{-1}(\sqrt{p})$ is the limiting function as $n \rightarrow \infty$.}
	\label{fig:figure1}
\end{figure}

Miller (1978) derived an inversion formula of the double arcsine tranformation as
\begin{equation}
p(\theta)=(1/2)\left[1-\mbox{sgn}(\cos(2\theta))\sqrt{1-[\sin(2\theta)+\{\sin(2\theta)-1/\sin(2\theta)\}/n^{\prime}]^2}\right],\label{eqn;3}
\end{equation}
where $n^{\prime}=n$ when samples are drawn from a population with the same sample size of $n$ and $n^{\prime}=k(\sum_{i=1}^k n_i^{-1})^{-1}$ when samples are drawn with different sample sizes of $n_i$ $(i=1,...,k$), where $k$ is the number of samples. Figure 2 plots the inverse function of the double arcsine tranformation (\ref{eqn;3}) for different $n^{\prime}$'s when $n^{\prime}=n$. Interestingly, even though its domain should be
\begin{equation}
\left[(1/2)\sin^{-1}\left(\sqrt{1/(n+1)}\right),\pi/4+(1/2)\sin^{-1}\left(\sqrt{n/(n+1)}\right)\right], \label{eqn;4}
\end{equation}
the inverse function is extended to exist outside the domain, showing an erratic behavior at both limits of its domain. For example, when $n=1$, the domain should be [0.392, 1.178] in Figure 2, but the inverse function still gives values outside that interval. As a remedy, therefore, for a small sample case the inverse function should be set to 0 below the lower limit of the domain and to 1 above the upper limit for one-to-one recovery of the original scale of proportion, i.e.
$$p(\theta)=0, \quad \mbox{if } \theta \leq (1/2)\sin^{-1}\left(\sqrt{1/(n+1)}\right),$$
and 
$$p(\theta)=1, \quad \mbox{if } \theta \geq \pi/4+(1/2)\sin^{-1}\left(\sqrt{n/(n+1)}\right).$$

Figure 2 also indicates that when studies with large sample sizes are included in the meta-analysis, the inverse of double arcsine tranformation can be approximated by its limiting function, i.e. $p(\theta)=\sin^2(\theta)$, $\theta \in [0,\pi/2]$, which is the inverse of the simple arcsine transformation function given in (\ref{eqn;2}).

\begin{figure}
	\centering
	\includegraphics[width=1\linewidth]{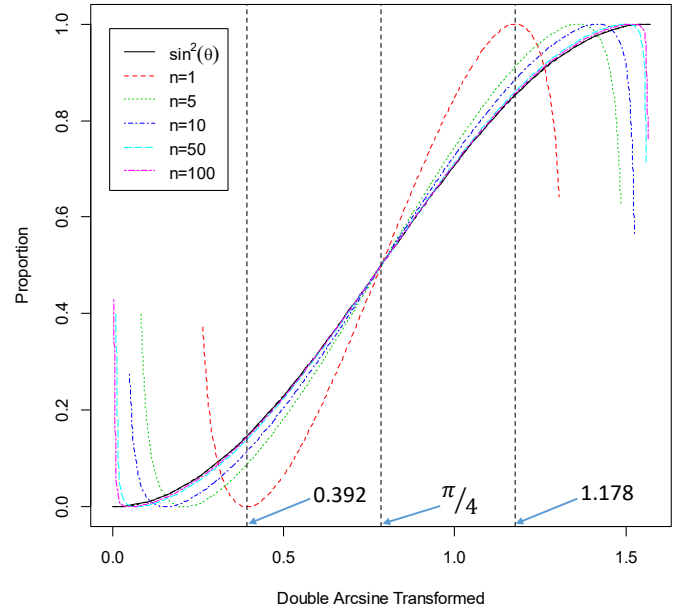}
	\caption{Graph of inverse double arcsine transformation for different sample sizes; $\sin^2(\theta)$ is the limiting function as $n \rightarrow \infty$.}
	\label{fig:figure2}
\end{figure}

\section{Accuracy of the large-sample approximation}
Figure 1 indicates that the accuracy of the approximation would depend on the true proportion $p$. Since the transformation is monotonely increasing and symmetric at the center, i.e. when $p=0.5$ and $\theta(0.5)=\pi/4$, we define the accuracy measure for the approximation as the maximum percent error (MPE),
\begin{equation}
\delta(p)=\sup_p \frac{|\sin^{-1}(\sqrt{p})-\theta(p)|}{\sin^{-1}(\sqrt{p})},\label{eqn;4.1}
\end{equation}
which measures the precent maximum difference between the double arcsine tranformation and its limiting function. The maximum occurs both at $p=0$ and $p=1$, but we will use the MPE at $p=1$ for mathematical convenience, i.e.
\begin{eqnarray}
\delta(1)&=&\frac{\sin^{-1}(\sqrt{1})-\theta(1)}{\sin^{-1}(\sqrt{1})} \nonumber \\
      &=&\frac{1}{2}-\frac{1}{\pi}\sin^{-1}\left(\sqrt{n/(n+1)}\right), \label{eqn;5}
\end{eqnarray}
  Since the MPE in (\ref{eqn;5}) is a function of the sample size $n$, it can be used to determine the sample size that would provide the MPE at a prespecified accuracy level $\epsilon$ by setting $\delta(1)=\epsilon$, where $\epsilon \in (0,1/2)$. After simple algebraic and trigonometric manipulation, we have the required sample size as a function of the accuracy level,
\begin{equation}
n=\tan^2\left(\pi\left(1/2-\epsilon\right)\right), \quad \epsilon \in (0,1/2). \label{eqn;6}
\end{equation}
Note that the sample size explodes to the infinity when the percent error $\epsilon=0$, as expected. For non-zero $\epsilon$  values, for example when $p=0.01$ and 0.05, the formula (\ref{eqn;6}) gives $n=1,013$ and 40, respectively, implying that the approximation will be at the accuracy level of 1\% and 5\% in terms of the MPE relative to the limiting function. Conversely, we can directly determine the MPE for a given sample size. For $n=200$ and 500, for example, the formula (\ref{eqn;5}) gives the MPE of 2.2\% and 1.4\%, respectively.

\section{Conclusion}
In this note, we presented details on the domain and range of the double arcsine tranformation, focusing on the domain of the inverse transformation. We noticed that the mathematical formula of the inverse function of the double arcsine tranformation should be used with caution due to its erratic behavior at both limits of its domain. The limiting function of the inverse tranformation reduces to the simple arcsine inverse tranformation, $\sin^2(x)$ ($x \in [0,\pi/2]$), which can be used for large sample approximation. For small sample cases, the inverse function should be set to 0 below the lower limit of the domain and to 1 above the upper limit to recover the original scale of the proportion. We also proposed a simple accuracy measure, the maximum percent error (MPE), of the large sample approximation, which can be used to determine the sample size that would provide a certain accuracy level, and conversely to determine the accuracy level of the approximation given a sample size.


\end{document}